\documentclass[journal=jctcce,manuscript=article]{achemso}

\usepackage{graphicx}% Include figure files
\usepackage{amsmath}
\usepackage{amssymb}
\usepackage[utf8]{inputenc}
\usepackage[T1]{fontenc}
\usepackage[colorlinks,citecolor=blue]{hyperref}

\newcommand{\vq}{q}

\title{
Fast and accurate multidimensional free energy integration
}

\author{J\'er\^ome H\'enin}
\email{jerome.henin@ibpc.fr}
\affiliation{CNRS, Université de Paris, UPR 9080, Laboratoire de Biochimie Théorique, Paris, France}
\affiliation{Institut de Biologie Physico-Chimique-Fondation Edmond de Rothschild, PSL Research University, Paris, France}

\date{\today}

\begin{document}

\begin{abstract}
Enhanced sampling and free energy calculation algorithms of the Thermodynamic Integration family (such as the Adaptive Biasing Force method, ABF) are not based on the direct computation of a free energy surface, but rather of its gradient. Integrating the free energy surface is non-trivial in dimension higher than one. Here the author introduces a flexible, portable implementation of a Poisson equation formalism to integrate free energy surfaces from estimated gradients in dimension 2 and 3, using any combination of periodic and non-periodic (Neumann) boundary conditions. The algorithm is implemented in portable C++, and provided as a standalone tool that can be used to integrate multidimensional gradient fields estimated on a grid using any algorithm, such as Umbrella Integration as a post-treatment of Umbrella Sampling simulations. It is also included in the implementation of ABF (and its extended-system variant eABF) in the Collective Variables Module, enabling the seamless computation of multidimensional free energy surfaces within ABF and eABF simulations. A Python-based analysis toolchain is provided to easily plot and analyze multidimensional ABF simulation results, including metrics to assess their convergence. The Poisson integration algorithm can also be used to perform Helmholtz decomposition of noisy gradients estimates on the fly, resulting in an efficient implementation of the projected ABF (pABF) method proposed by Lelièvre and co-workers. In numerical tests, pABF is found to lead to faster convergence with respect to ABF in simple cases of low intrinsic dimension, but seems detrimental to convergence in a more realistic case involving degenerate coordinates and hidden barriers, due to slower exploration. This suggests that variance reduction schemes do not always yield convergence improvements when applied to enhanced sampling methods.
\end{abstract}

% \pacs{}% insert suggested PACS numbers in braces on next line
% {\bfseries Keywords: molecular dynamics, free energy, enhanced sampling, ABF}

\maketitle %\maketitle must follow title, authors, abstract and \pacs

% 
% {\itshape
% Reviewers: Fabio Pietrucci (cite), Mahmoud Moradi (cite), Jeff Comer (cited),
% Cameron Abrams (cite Testing Convergence of Different Free-Energy Methods in a Simple Analytical System with Hidden Barriers )
% Régis Pomès ?
% Olgun Guvench (? user)
% }

% \tableofcontents

\section{Introduction}

In molecular dynamics simulations, physical interactions intervene solely in the form of forces,
which is why empirical potential energy functions are aptly nicknamed ``force fields''.
Energy estimation is, in that sense, a secondary concern.
For the purpose of enhanced sampling, biased dynamics can be obtained through biasing forces, which may or may not derive from a biasing potential.
Whereas Adaptive Biasing Potential methods\cite{Huber1994, Laio2002, Marsili2006, Babin2008, Dickson2010} involve the estimation of a free-energy surface (FES), Adaptive Biasing Force (ABF) methods\cite{Darve2001} are built around the estimation of the free energy gradient, in the spirit of Thermodynamic Integration (TI).\cite{Kirkwood1935}
A key difference between those two perspectives is that the free energy gradient is a local, absolute quantity, whereas free energy measurements are relative and hence non-local.
Due to the local character of ABF and the absence of ``fill rate'' measuring the steady-state work exerted by the bias, the simulation time necessary to cross a barrier is independent from its height. 

Usually, however, the quantity of interest is not the free energy gradient, but the free energy surface itself.
In addition, the mean force used to estimate this quantity in multidimensional ABF\cite{Darve2008, Henin2010a} is not exactly the gradient of any scalar potential, although it converges towards such a gradient.
This makes estimating the free energy a non-trivial problem.

In this contribution I present the implementation of a solution to this problem based on original work by Lelièvre and coworkers.\cite{Lelievre2010, Alrachid2015}.
The implementation is part of the Collective Variables Module\cite{Fiorin2013}, 
The formalism expresses the free energy surface as solution to a Poisson problem, making it amenable to robust numerical schemes.
I also present a ``real-world'' implementation of projected ABF (pABF) proposed by the same authors, and use it to assess its practical merits when sampling biomolecular systems. 
The results are found to vary greatly between a low-dimension model and a more realistic system exhibiting many coupled slow degrees of freedom, in a similar fashion to large biomolecular assemblies.

\section{Theory and methods}

Consider a classical-mechanical system with Cartesian coordinates $q$, subject to the potential energy function $V(\vq)$, and evolving  under Langevin dynamics or another type of thermostatted classical dynamics that samples from the canonical ensemble at temperature $T$, or inverse temperature $\beta \equiv 1/(k_B T)$.
That is, the limiting distribution of the dynamics is the Boltzmann distribution $\rho(q) \propto \exp(-\beta V(\vq))$.
In the following we will focus on configurational statistics, rather than dynamical properties.
Given an arbitrary set of collective variables $z = \xi(q)$, the free energy surface associated to $z$ is defined, up to an additive constant, by:
\begin{equation}
\label{eq:fes}
  A(z) \equiv  - kT \ln  \int e^{-\beta V(\vq))} \delta(\xi(\vq) - z) d\vq  
\end{equation}
That is, $A(z) =  - kT \ln \rho(z)$, where $\rho(z)$ is the marginal probability density of variable $z = \xi(q)$.

\subsection{ABF yields estimates of the free energy gradient}

ABF was originally described and implemented for a unidimensional collective variable.\cite{Darve2001,Henin2004}
ABF in dimension greater than one\cite{Darve2008, Henin2010a} is built around an estimator of the free energy gradient along the chosen collective variables.
The implementation of ABF within the Colvars Module\cite{Fiorin2013} implements a formalism derived by Ciccotti et al.\cite{Ciccotti2005}, based on projection of atomic forces in Cartesian coordinates onto the collective variables.
The gradient of $A$ may be estimated as:
\begin{equation}
\label{eq:gradient}
G(z) = \left\langle \nabla_q V(q) \cdot \mathbf{v}(q) - \frac{1}{\beta} \nabla_q \cdot \mathbf{v}(q) \right\rangle_z
\end{equation}
where $\mathbf{v}(q)$ is an arbitrary vector field satisfying certain orthonormality conditions with the collective variable gradient $\nabla_q \xi$, and the brackets indicate a canonical average conditioned by $\xi(q) = z$.\cite{Ciccotti2005,Henin2010a}
Empirically, Thermodynamic Integration has been found to be more efficient than other free energy estimators, regardless of the method used to perform sampling;\cite{Fiorin2013, Cuendet2014, Mones2016}
yet no theoretical justification has been put forward.

%% eABF
If force projection is not practical, the extended-system ABF (eABF) approach can be used, which estimates the gradient of a biased free energy surface $A^k$:\cite{Lelievre2007,Lesage2017}
\begin{equation}
\label{eq:eabf_fes}
  A^k(\lambda) \equiv  - kT \ln  \int e^{-\beta V(\vq)}  e^{-\frac{\beta k}{2}   (\xi(\vq) - \lambda)^2} d\vq
\end{equation}
Note that Equation~\ref{eq:eabf_fes} is related to Equation~\ref{eq:fes} simply by substitution of the Dirac distribution with a Gaussian kernel, of variance inversely proportional to the coupling constant $k$.
eABF can be complemented with unbiased estimators of the free energy gradient $\nabla A(z)$, such as Corrected z-averaged restraint (CZAR)\cite{Lesage2017} or Umbrella Integration.\cite{Kastner2005,Zheng2009,Fu2016}

% eABF : Convolved PMF. Parameters: coupling width and time constant (unphysical).
% Convergence of an ABF simulation requires exploration of collective variable space, overcoming barriers along that manifold, but also sampling of the other degrees of freedom (the ``orthogonal space''), overcoming any barriers along those directions.
% In the presence of such hidden barriers, the choice of collective variables is crucial to the convergence of adaptive sampling methods.\cite{Paz2018}
% Some approaches have deen devised to explicitly enhance sampling in a direction corresponding to hidden barriers.
% Another approach is, rather than overcome hidden barriers, to \textit{bypass} them.
% That is made possible by sufficiently long simulations where diffusion is enhanced along the colvars, allowing for the same region of colvar space to be visited multiple times, potentially along different channels.
% A more direct approach to sampling a variety of channels is the multiple-walker strategy, where several simulations proceed separately, with possible coupling between them by sharing information.

The initial implementation of ABF in the Colvars Module\cite{Henin2010a} only yielded free energy gradient estimates: multidimensional free energy surfaces were integrated using a separate tool (\texttt{abf\_integrate}), which performs discrete Markov-Chain Monte-Carlo (MCMC) sampling with a metadynamics-like history-dependent bias.
That method is algorithmically simple, general, and importantly, implemented in arbitrary dimension.
However, its convergence properties are unfavorable and difficult to tune.
The estimate tends to fluctuate around the optimal value of the free energy, which generally results in noisy estimates, and the bias increments can be made to decay to zero over time, but this introduces an additional parameter.
The present, fully deterministic approach solves these issues, improving both the performance and reliability of the integration, and simplifying its use by eliminating most tunable parameters.

\subsection{Free energy surface as ``integral'' of a non-conservative vector fields}

Considering a \textbf{multidimensional} collective variable $z = \xi(q)$, we note  $G_t(z)$ the estimate at time $t$ of the free energy gradient.
Obtaining the corresponding FES $A_t(z)$ is not as simple a task as integrating a scalar function.
The main reason for that is that for finite $t$,  $G_t$ is not a gradient.
Due to statistical noise, it is not conservative, meaning that its integral over closed curves is not zero, and as a result it does not derive from any potential (scalar field).
Therefore the intuitive differential equation

\begin{equation}
\nabla A_t = G_t
\end{equation}

admits no solution for $A_t$.
Therefore, the question ``\textit{What FES is $G_t$ the gradient of?}'' is not a well-posed problem.

\subsubsection{Poisson integration}
\label{sec:Poisson}

Lelièvre and coworkers have proposed\cite{Lelievre2010, Alrachid2015} to express the free energy as the solution to a well-posed Poisson problem, by noting that the divergence of $G_t$ is an estimator of the Laplacian of the free energy\cite{Alrachid2015}:

\begin{equation}
\nabla^2 A_t = \nabla \cdot G_t
\label{eq:poisson}
\end{equation}

The present algorithm consists in a finite-difference solution of Equation~\ref{eq:poisson} on a regular grid.
The solution $A_t$ minimizes the $L^2$ distance $||\nabla A_t - G_t||_2$\cite{Alrachid2015}, that is, it answers the question ``\textit{What FES has a gradient that is as close as possible to $G_t$?}'' (in the least-squares sense).
Equation~\ref{eq:poisson} only defines $A_t$ up to a linear function, so it needs to be complemented by appropriate boundary conditions, as discussed below.

Other methods to reliably reconstruct a free energy surface from gradient estimates differ by the basis functions used and the optimization method.
In particular, the minimization of $||\nabla A_t - G_t||_2$ can be expressed directly as a set of linear equations involving the parameters of the free energy surface.\cite{Kastner2009} This is used in the On-The-Fly Parameterization (OFTP) method,\cite{Abrams2012} which has been implemented up to dimension 2 on a basis of \textit{chapeau} functions.\cite{Paz2018a}
Expressing the minimization problem as stated in OTFP in our numerical context leads to the same expressions as when starting from the Poisson problem.
Maragliano and Vanden-Eijnden have proposed to use a variational optimization on a basis of radial functions.\cite{Maragliano2008}
A Bayesian strategy that integrates any knowledge about the error in gradient data is to apply Gaussian Process Regression\cite{Stecher2014}, which has been coupled with ABF\cite{Mones2016}.
The present formulation of finite-difference Poisson integration is ideally suited to gradient data from ABF, which is estimated on a regularly-spaced dense grid.

\subsubsection{Finite-difference Poisson integration algorithm}

Integrating Equation~\ref{eq:poisson} requires two numerical steps: (1) compute the discretized divergence of the current gradient estimate $G_t$, completed with relevant boundary conditions, and (2) solve the resulting discrete Poisson problem using a Conjugate Gradients (CG) linear solver.\cite{Shewchuk1994, NumericalRecipes2002}
The conjugate gradients method is particularly efficient for sparse, symmetric, positive-definite linear systems, which is the case of the discrete Poisson equation.
In CG, the target vector is optimized in steps along a succession of mutually orthogonal search directions. A complete yet accessible (even entertaining) introduction to CG can be found online, in a little gem of a paper by Jonathan Shewchuck.\cite{Shewchuk1994}. 
The version of CG used here does not involve preconditioning.
Tests using the universal but simple-minded Jacobi preconditioning yielded no improvements.
While preconditioning is generally advised to improve the convergence of CG, it is still a subject of active research in the context of the Poisson equation with Neumann BCs\cite{Lee2021}, let alone mixed boundary conditions.

\begin{figure}[ht!]
 \centering
 \includegraphics[width=7cm]{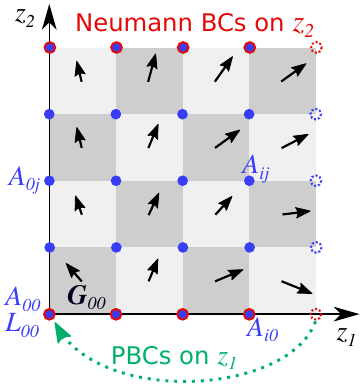}
 \caption{Two types of grids used to represent the free energy surface and its derivatives. The inner grid (shaded squares) is used for the discretized free energy gradient $\mathbf{G}_{ij}$ (arrows), averaged over bins. The outer grid (blue dots) is used for both the discretized free energy $A_{ij}$ and Laplacian $L_{ij}$. This example shows mixed boundary conditions: periodic along $z_1$, and non-periodic (Neumann) along $z_2$.}
 \label{fig:grids}
\end{figure}

In the Colvars implementation of ABF and eABF-CZAR,  the conditional average used to estimate the free energy gradient $G_t$ is collected within discrete bins, therefore the estimate is really the average gradient within the bin boundaries.
In 1D, integration is performed using a simple middle Riemann sum, yielding free energy values at the bin edges.
Because the gradient is a bin average accounting for all gradient values within the bin interval (Figure~\ref{fig:grids}), this is the most accurate scheme.
This is in contrast to purely local estimates of the gradient that could be obtained from constrained simulations; in that case, a higher-order integration scheme would be warranted to account for intermediate values of the gradient.

In accordance with that strategy, in the present scheme the free energy and its Laplacian are described on the same grid (blue dots in Figure~\ref{fig:grids}), whereas the gradient grid is shifted by half a grid interval in each dimension (shaded squares and arrows in Figure~\ref{fig:grids}):
\begin{align}
 z_i^k &= z_i^0 + k \delta z_i \\
 z_i^{\text{grad},k} &=  z_i^0 + \left(k+\frac{1}{2}\right ) \delta z_i 
\end{align}

As a result, the FES grid is larger than the gradient grid by one bin, except in PBC, where the FES grid wraps around and the last point is represented by the first point.
The Laplacian grid always has the same size as the FES grid.
This way, numerical differentiation can be performed as a centered two-point difference (or $2^d$-point in dimension $d$), and the discretized Laplacian takes a classic form, using a five-point stencil in 2D and a seven-point stencil in 3D.

\begin{equation}
 L_{ij} = \frac{A_{i-1,j} + A_{i+1,j} - 2 A_{i,j}}{\delta z_1^2} + \frac{A_{i,j-1} + A_{i,j+1} - 2 A_{i,j}}{\delta z_2^2}
\end{equation}

Periodic boundaries are specified by wrapping the discrete Laplacian stencil around the boundaries.
In non-periodic cases, Neumann boundary conditions are applied, that is, the value of the gradient in the normal direction to the boundary is imposed.
This choice is natural given the data, which specify the gradient values everywhere (including the normal gradient at the boundaries).
In contrast, classic boundary conditions such as Dirichlet cannot be applied here because there is no way to specify a priori any boundary values of the free energy itself.
In that non-periodic case, the boundary element of the ``Laplacian'' matrix is written, taking as an example the $z_2 = z_2^0$ boundary:
\begin{equation}
 L_{i0} =  \frac{1}{2} \; \frac{A_{i-1, 0} + A_{i+1, 0} - 2 A_{i,0}}{\delta z_1^2}
 +  \frac{1}{\delta z_2} \; \frac{A_{i, 1} - A_{i, 0}}{\delta z_2}
\end{equation}

Note that the $z_1$ term of the Laplacian is halved on the $z_2$ edge, and the $z_2$ term is replaced by the finite-difference normal gradient, \textit{further divided by} $\delta z_2$. These two modifications combined preserve the symmetry of the discrete Laplacian.\cite{Thomas2013, Chen2019}
Recall that the free energy and laplacian grids are each treated as a vector, so using the present notations, the discrete Laplacian operator is symmetric iff the coefficient of $A_{ij}$ in $L_{kl}$ is the same as that of $A_{kl}$ in $L_{ij}$ for all $(i, j)$ and $(k, l)$.

Semi-periodic conditions may occur if not all colvars are periodic. 
They are treated consistently with the two cases above, again, keeping the discrete Laplacian matrix symmetric.

Equation~\ref{eq:poisson} is solved using a conjugate gradients algorithm, taking into account the symmetry of the discrete Laplacian.
This only requires the Laplacian to be expressed as a left-multiply function, which applies the Laplacian operator to a given vector.
As a result there is no need to explicitly store the discrete Laplacian matrix, which is very sparse and of high dimension especially in the 3D case (dimension $(l \times m \times n)^2$, where $l$, $m$, $n$ are the grid sizes).
Therefore an efficient implementation can be written without dependence on sparse matrix storage and manipulation algorithms.

\subsubsection{Practical use of Poisson integration in the Colvars Module}

Within the ABF\cite{Henin2010a} and eABF-CZAR\cite{Lesage2017} algorithms in Colvars, integration of 2D and 3D FES is enabled by default.
This has been included in the Colvars Module since version 2017-12-14, in NAMD 2.13 and LAMMPS verison stable\_16Mar2018.
It is controlled by the \texttt{integrateTol} and \texttt{integrateMaxIterations} keywords, which set the tolerance and maximum number of iterations of the conjugate gradients solver.

Poisson integration routines are implemented within the generic \texttt{colvargrid} class of the Colvars Module, used to handle quantities discretized on regular grids. Thus they can be re-used by any algorithm that follows a Thermodynamic Integration-like formalism to estimate a multidimensional free energy gradient, such as Umbrella Integration\cite{Kastner2005} or the combination of metadynamics and Thermodynamic Integration.\cite{Fiorin2013}
In particular, I provide a lightweight, standalone command-line tool \texttt{poisson\_integrator} to parse gradient grid files and perform the integration.
It can be used to integrate any gridded gradient estimate obtained using unrelated algorithms.

In particular, the ABF/Colvars implementation can be used to merge data from independent simulations covering different regions of colvar space (\textit{stratification}), which may be overlapping or simply adjacent.
This is done by providing data from several previous ABF simulations as input (using the \texttt{inputPrefix} keyword), and running a single timestep of simulation to trigger an ABF update.
In that case, gradient estimates from all regions are combined by ABF, then integrated using the present algorithm, which results in a single integrated FES covering the complete domain of colvars.

% Merging stratified CZAR is fixed in Colvars in new_abf branch 
% In the special case of stratified eABF/CZAR, CZAR gradient files should be concatenated and passed to the \texttt{poisson\_integrator} tool.
% This avoids concatenating histogram files from different strata, which are not matched at the boundaries, and would introduce discontinuities in the CZAR gradient estimate.

\subsection{Projected ABF for variance reduction in real-world simulations}

\subsubsection{Helmholtz projection of the numerical mean force}

Alrachid and Lelièvre proposed\cite{Alrachid2015} to use the numerical gradient of the estimated FES $A_t$, $\nabla A_t$, as the basis for the biasing force, thus performing a Helmholtz projection of $G_t$ onto the space of irrotational vector fields, and eliminating its non-conservative components.
Thus the ``mean force'' estimate of the gradient is replaced with the closest quantity (in the least squares sense, see section~\ref{sec:Poisson}) that behaves like a true gradient, i.e. a conservative vector field.
The spurious degrees of freedom present in the mean force are eliminated when computing the discrete Laplacian as part of the Poisson problem.

This pABF approach is an instance of ABF with kernel-based force reconstruction, like ABF(GPR),\cite{Mones2016} where the free energy surface is reconstructed via Gaussian Process Regression.\cite{Stecher2014}
Using a proof-of-concept implementation, Alrachid and Lelièvre performed numerical experiments of pABF on a toy model: repulsive particles in dimension two, with a double-well potential on a single degree of freedom.
This demonstrated reduced variance of the biasing force in pABF, resulting in faster convergence than standard ABF dynamics.
However, the implementation did not support molecular simulations, and consequently, did not allow for numerical tests of pABF on more realistic and challenging systems.

\subsubsection{Optimized on-the-fly integration for pABF}

The pABF implementation of Colvars relies on Poisson integration as described above.
The integrated FES is updated with at finite time intervals, and the biasing force is calculated as a finite-difference gradient of the latest integrated FES.
Since the gradient and FES grids are shifted by half a bin width, the numerical differentiation is a $2^d$-point centered finite difference.

The Poisson integration algorithm as described above is well-suited for a single, isolated free energy estimation.
However, for pABF in particular, the FES estimate must be updated many times at short time intervals.
Furthermore, the values of $G_t$ vary more and more slowly as the algorithm approaches convergence; then, most of the work in successive updates becomes redundant.
In pABF/Colvars, this redundant work is avoided in two ways: the divergence $\nabla \cdot G_t$ is updated locally where new samples have been collected. Then, at pABF update times, the conjugate gradients optimizer is restarted from the previous value of $A_t$, leading to rapid convergence.
Finally, the optimization is performed alternately at two different precisions: at high precision when (re-)initializing the algorithm or outputting the FES, and at a lower precision upon frequent updates for pABF.

Accordingly, pABF in Colvars is controlled by the following user parameters:\\
\texttt{pABFIntegrateFrequency}, \texttt{pABFintegrateTol}, and \texttt{pABFintegrateMaxIterations}.

\section{Numerical methods}

Numerical tests of the Poisson integration routines shown in Figures~\ref{fig:Poisson}, \ref{fig:Poisson3d} are based on the following model free energy surfaces:
\begin{align}
%   p = sin(x*deg2rad)*-.5*cos(2*y*deg2rad) + .5
A^{2d} &= \frac{1}{2} (\sin(x) \cos(2y) + 1) \label{eq:model2d} \\  
A^{3d} &= \frac{1}{2} (\sin(x) \cos(2y) + 1) + \cos(z) + 1
\label{eq:model3d}
\end{align}
The gradients of those surfaces were calculated on regular grids by Python scripts, and fed to the Poisson integration algorithm.
The resulting integrated ``free energy'' surfaces were then compared to their exact values.

Convergence of the integration was monitored using the relative error between the computed discrete Laplacian and its target value  $||\nabla^2 A - \nabla \cdot G||_2 / ||\nabla \cdot G||_2 $ (Equation~\ref{eq:poisson}), and the root-mean-square deviation (RMSD) between the integrated FES and the exact FES (Equations~\ref{eq:model2d} and \ref{eq:model3d}).
The relative error is always available and is the stopping criterion of the conjugate gradients solver, whereas the RMSD to the exact free energy RMSD is only available for synthetic gradient data.

The test systems considered for ABF and pABF simulations were two small peptides in vacuum that were used in many studies before: the dipeptide mimic N-acetyl-N-methyl-L-alanylamide (NANMA), and the helical peptide deca-alanine, which exhibits a complex conformational landscape.\cite{Henin2010a}

Simulations were run using NAMD~2.14~\cite{Phillips2020} coupled with the current version of the Collective Variables Module~\cite{Fiorin2013}.
In vacuo systems were modeled by the CHARMM22 force field, and simulated with a base timestep of 0.5~fs and outer timesteps of 2 and 4 femtoseconds for non-bonded and long-range forces respectively. Langevin dynamics at 300~K was run with a damping coefficient of 5~ps$^{-1}$.

In the NANMA example, ABF dynamics was applied to the Ramachandran angles $\varphi$ and $\psi$, with a bin width of 5 degrees. The \texttt{fullSamples} parameter was set to 20 time steps.

To compare standard ABF followed by Poisson integration with a FES obtained directly, well-tempered metadynamics simulation\cite{Barducci2008} was run on the same coordinates using the Colvars Module.
The metadynamics hill height was 0.1~kcal/mol, the hill width was 9 degrees, and the hills were added every 50 timesteps, with a bias temperature of 6000~K. The biasing potential was projected onto a grid with 5-degree spacing to allow for direct comparison with ABF/Poisson.
20 replicas were run for 100~ns each, and the resulting FES were averaged to give the comparison FES.
Similarly, the free energy gradient from 20 independent ABF simulations were averaged, then integrated to give the ABF FES shown in Figure~\ref{fig:Poisson} (right panel).

ABF simulations on deca-alanine use dihedral PCA~\cite{Altis2007} (dPCA) modes as collective variables. dPCA was performed using the software Carma~\cite{Glykos2006} on a dataset of 7 structures, representing the metastable conformations identified in reference~\citenum{Henin2010a}.
\texttt{fullSamples} was set to 200.
In all pABF simulations, the integrated free energy was updated every 200 steps, with a tolerance of $10^{-4}$ and a maximum of 100 conjugate gradient steps.
In simulations of larger, solvated systems, the integration cost would be comparatively negligible.

Convergence of free energy gradient estimators was assessed as follows.
For each set of parameters, ten replica simulations were started using pseudo-random initial velocities and different seeds for the pseudo-random generator providing Langevin stochastic forces.
Values of the estimators to be tested were saved regularly using the \texttt{historyFreq} parameter of ABF/Colvars.

The average of final estimates of the free energy gradient from all replicas was used as reference for measuring convergence.
The gradient and free energy estimates of each replica were compared to this reference using an RMSD distance.
The average and standard deviations of these distances over the ten replicas were used to assess the accuracy and rate of convergence of estimators, and are plotted in Figures~\ref{fig:pabf_nanma} and \ref{fig:pabf_10ala}.

% \newpage
\section{Results}

\subsection{Robust and accurate free energy landscapes in dimension 2 and 3}

\begin{figure}[!ht]
\centering
 \includegraphics[width=0.49\textwidth]{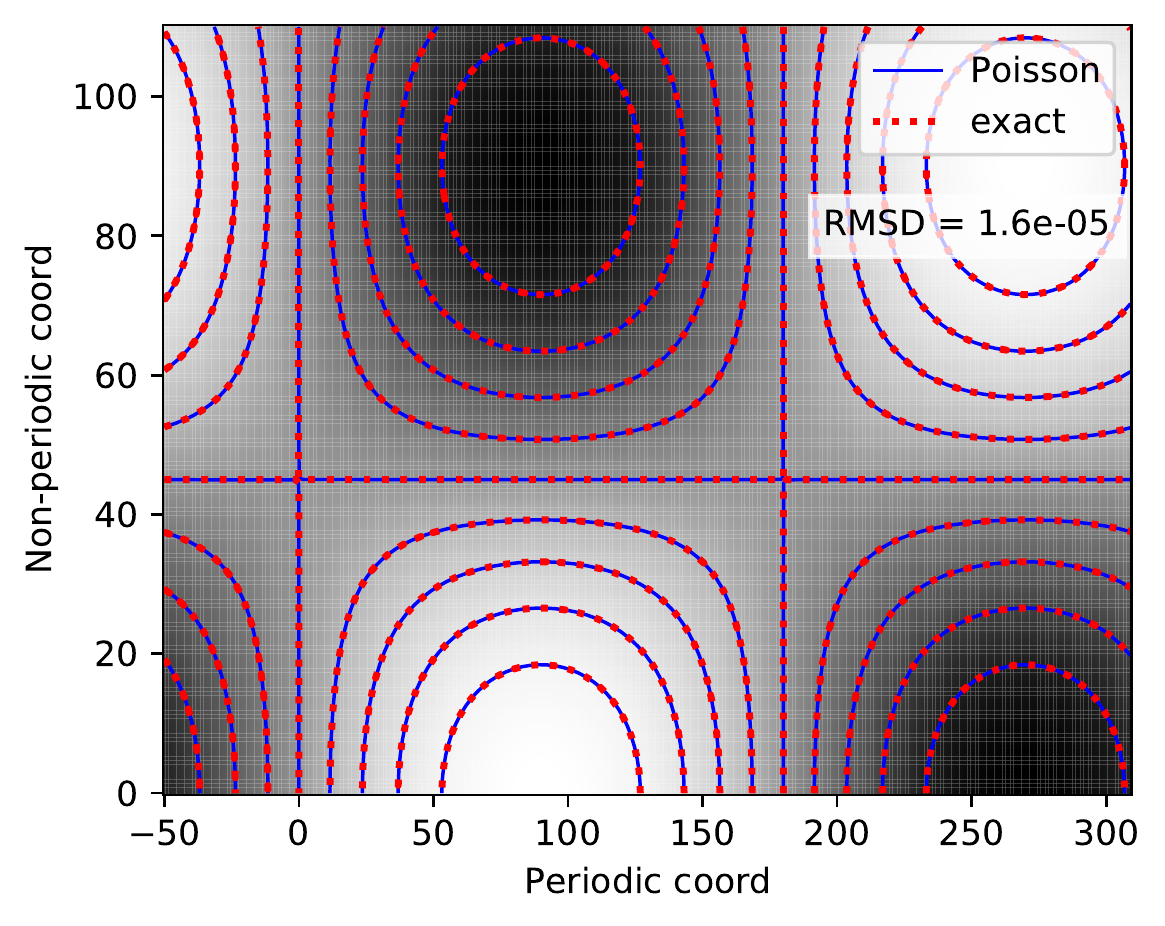}
 \includegraphics[width=0.49\textwidth]{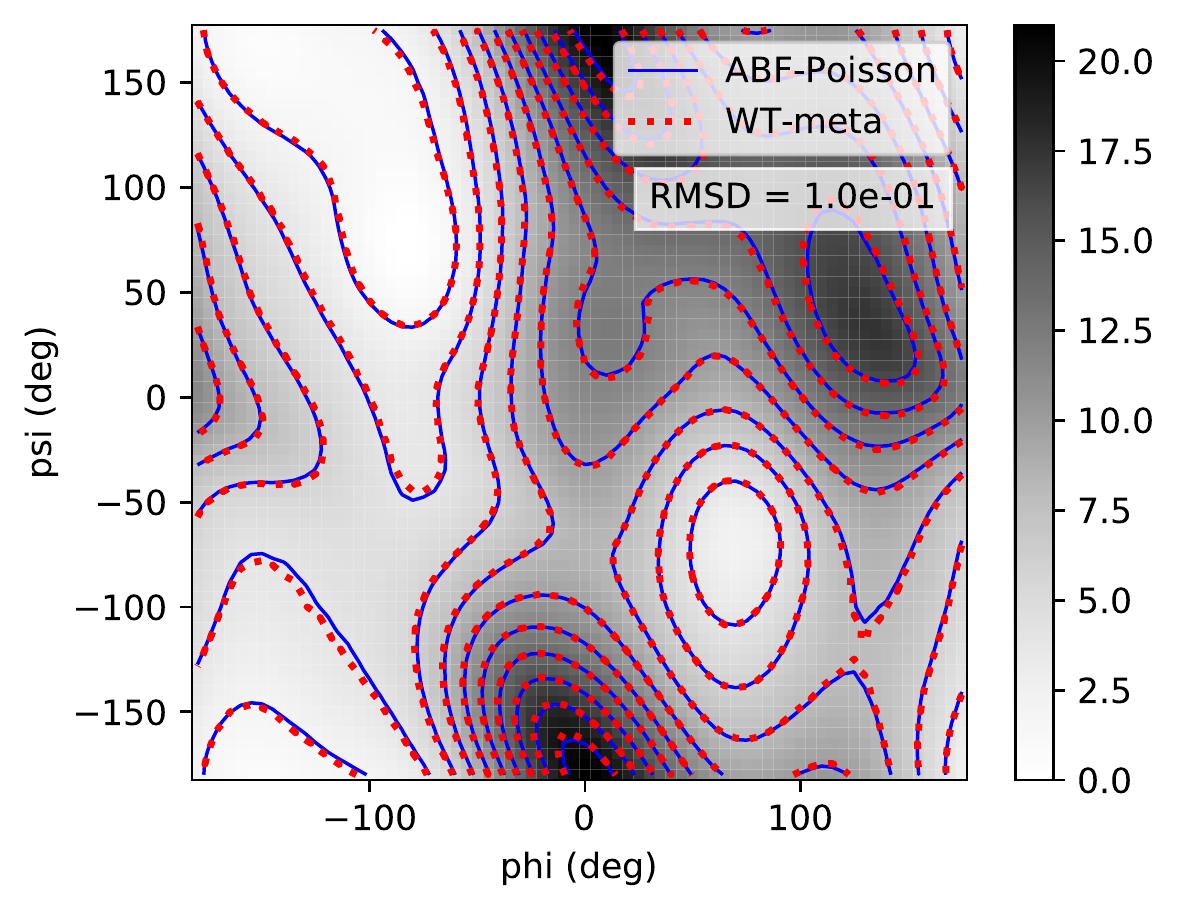}

 \caption{Left: Integration of a model 2d free energy surface (Eq.~\ref{eq:model2d}) from its discretized gradient in semi-periodic boundary conditions, using a discrete Poisson equation.
Level lines of the exact free energy surface (blue, solid) are superimposed with those of the result of Poisson integration (red, dotted). Free energy values are given by the grayscale background. The first coordinate is periodic, whereas the second is not. The free energy scale is 0 to 1.
Right: Comparison of the level lines for the free energy surface of NANMA, from ABF simulations with Poisson integration and from independent well-tempered metadynamics simulations. The free energy is given by the color scale (unit: kcal/mol).
}
\label{fig:Poisson} 
\end{figure}

\begin{figure}[!ht]
\centering
 \includegraphics[width=0.49\textwidth]{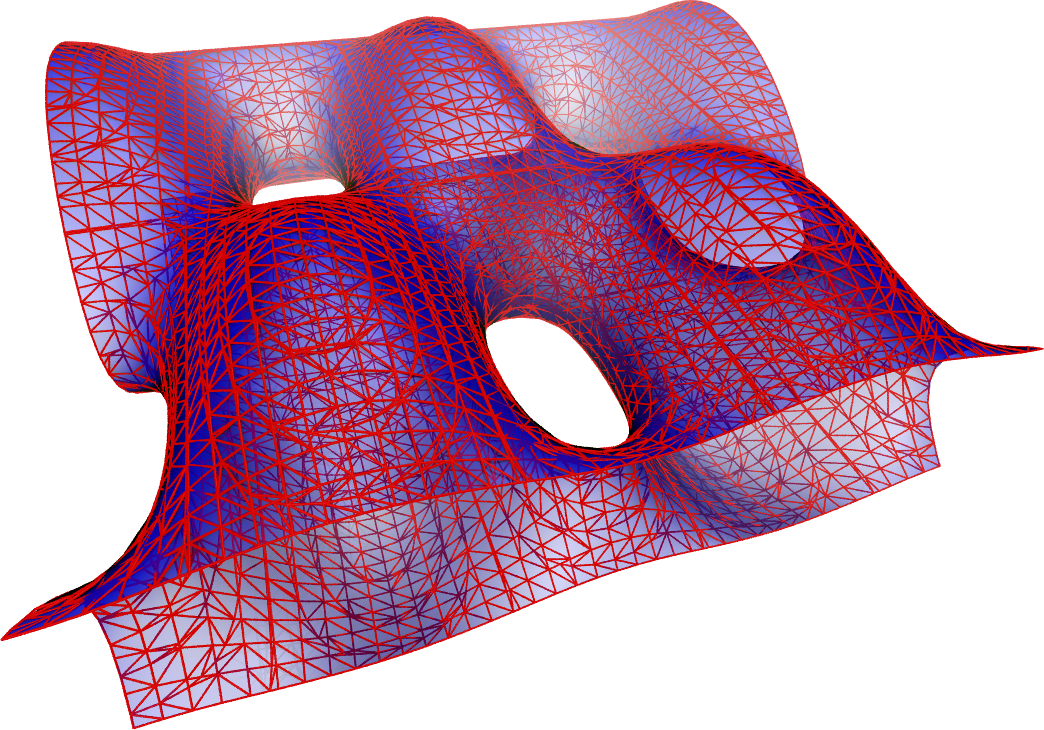}
 \includegraphics[width=0.49\textwidth]{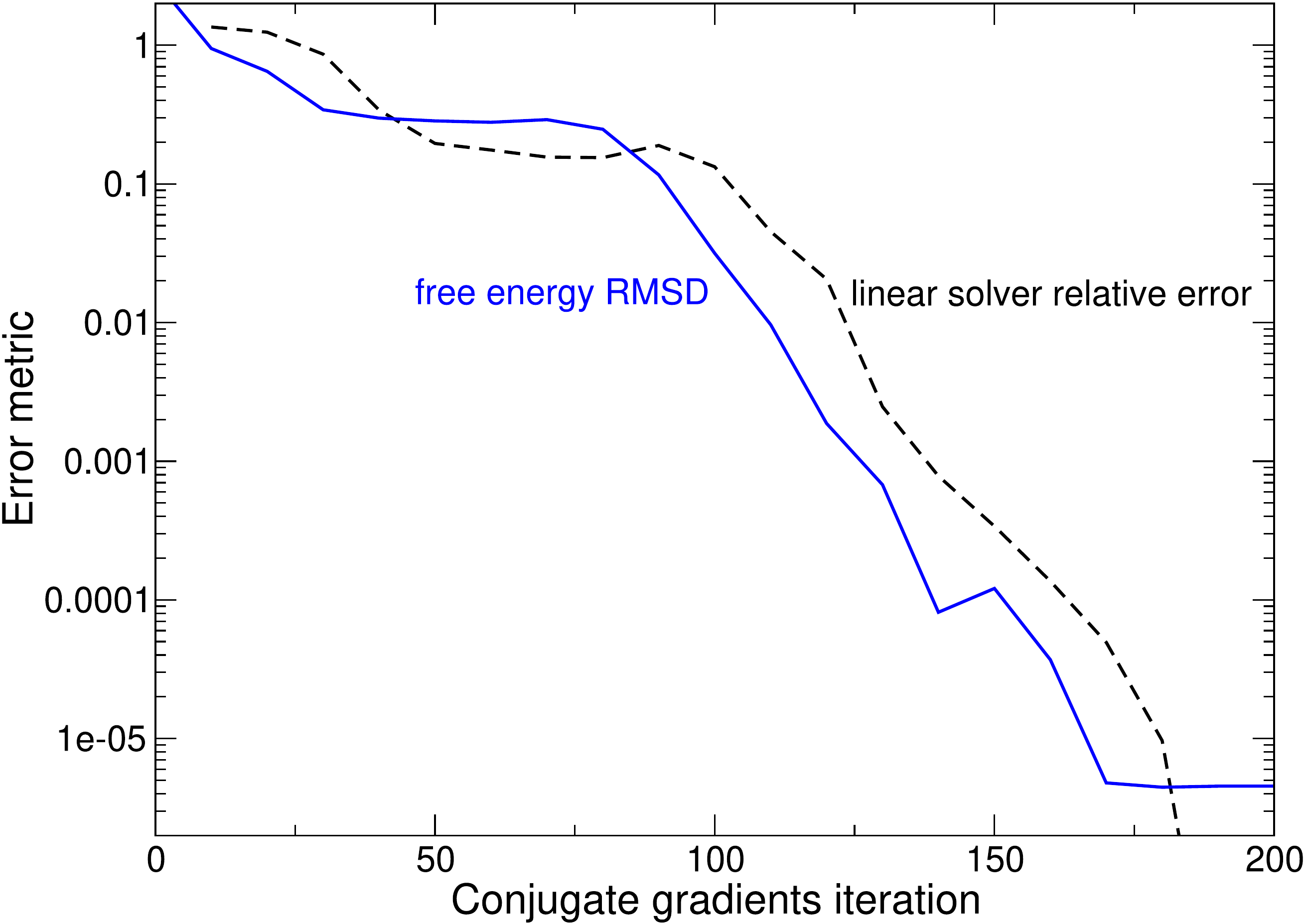}

\caption{Left: Poisson integration of 3D gradients in semi-periodic boundary conditions. Isosurface of a model 3d free energy landscape (Eq.~\ref{eq:model3d}), solid blue) and the same isosurface for the potential integrated from its discretized gradient (red mesh). The two horizontal coordinates are periodic, while the vertical one follows Neumann boundary conditions. Graphics rendered using VMD.\cite{Humphrey1996}
Right: Convergence of Poisson integration by conjugate gradients. The relative error of the linear solver is represented by the dashed black line, and the free energy RMSD from the reference is shown as a solid blue line.
 }
\label{fig:Poisson3d}
\end{figure}

The left panel of Figure~\ref{fig:Poisson} illustrates the accuracy of Poisson integration on an analytical 2d potential, in a non-trivial case of mixed boundary conditions: periodic boundary conditions on the first coordinate, and Neumann boundary conditions on the second.
The 3d FES defined by Eq~\ref{eq:model3d}, also in semi-periodic boundary conditions, is represented in Figure~\ref{fig:Poisson3d} together with the convergence graph.
As the conjugate gradients solver converges, the error between the integrated FES and the exact expression reaches a plateau at $5\times 10^{-6}$, which is the residual due to the discretization error.

The right panel of Figure~\ref{fig:Poisson} illustrates the accuracy of the integrated FES on a real rather than synthetic data: the 2d FES of NANMA, with a reference surface computed using well-tempered metadynamics simulations. The RMSD between the two surfaces is only 0.1~kcal/mol for free energy values ranging from 0 to 22~kcal/mol, and can be ascribed to residual statistical variance in both free energy estimates, as well as discretization error in ABF and the finite width of Gaussian kernels in metadynamics.

Free energy gradients in dimension higher than 3, which are rarely used, can still be integrated using the existing MCMC integration tool in arbitrary dimension.\cite{Henin2010a}

\subsection{Noise reduction by gradient projection in pABF}

\begin{figure*}[!ht]
\centering
 \includegraphics[height=0.7\textheight]{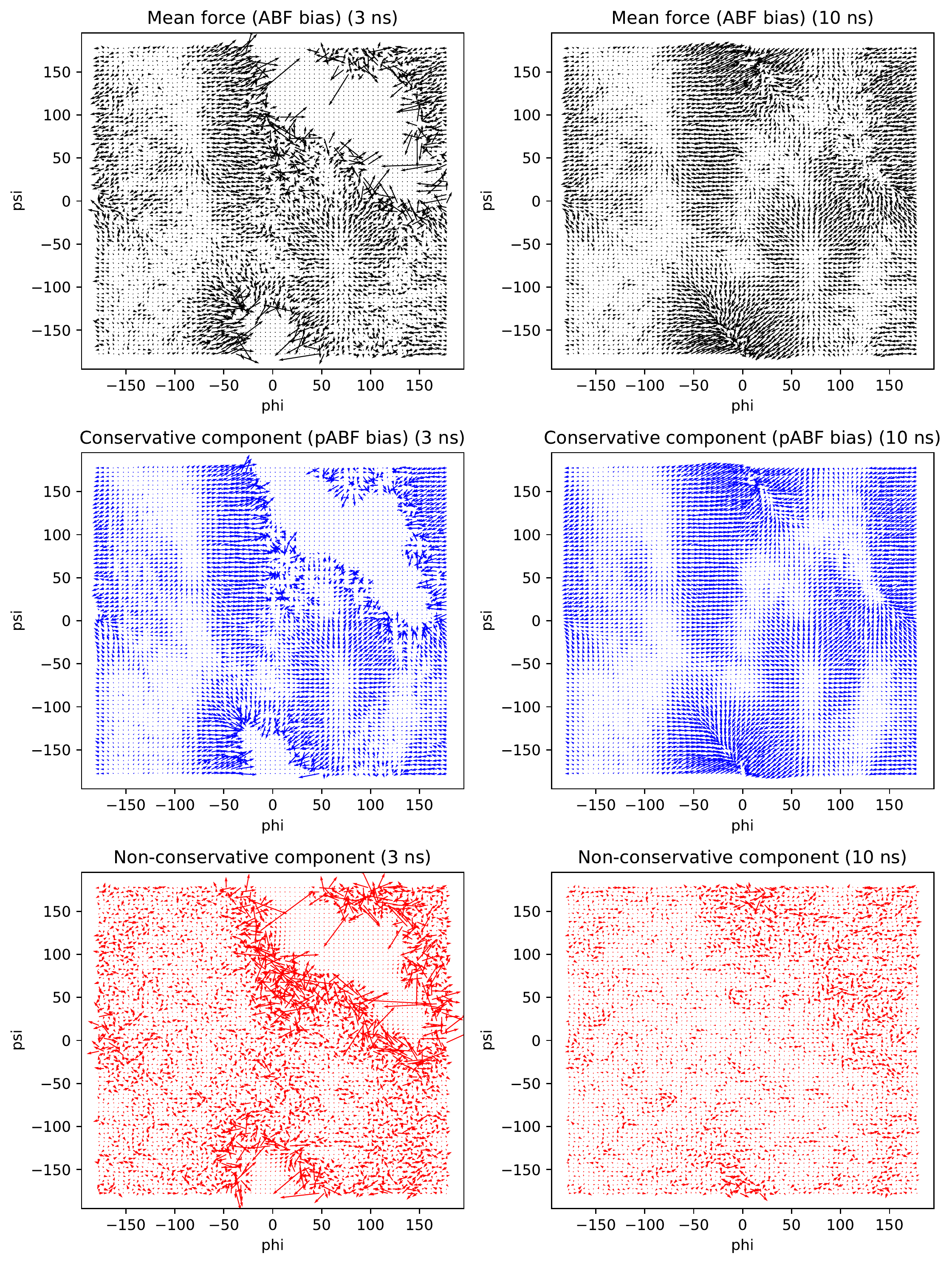}

 \caption{Helmholtz decomposition of a noisy discrete estimate of the free energy gradient for the Ramachandran angles of NANMA, 
 in an ABF simulation after 3 ns (left column) and 10 ns (right column) of sampling.
 Top (black): mean force estimate $G_t$ of the free energy gradient from ABF simulations.
 Middle (blue): finite-difference gradient $\nabla A_t$ of the free energy surface obtained from $G_t$ by Poisson integration.
 Bottom (red): residual $G_t - \nabla A_t$ representing the non-conservative component of the gradient estimate. Arrow scale is twice that of the upper panels to enhance the visibility of small vectors.}
\label{fig:Helmholtz}
\end{figure*}

Figure~\ref{fig:Helmholtz} illustrates the Helmholtz decomposition of noisy gradient data from an ABF simulation of the NANMA system.
The top left panel shows the estimate of the gradient $G_t$ from a short, unconverged ABF simulation.
The middle left panel is the conservative component of $G_t$. It has zero curl, and its divergence is the Laplacian of the free energy.
This vector field preserves the features of $G_t$ corresponding to variations of the underlying free energy, but is smoother.
The difference $G_t - \nabla A_t$ (bottom left) is a purely non-conservative field with zero divergence and nonzero curl.
It appears very noisy, and the only discernible features are loops, which attest to its non-conservative character.
Taken together, these illustrate that the conservative component $\nabla A_t$ is a reduced-variance estimate of the free energy gradient, compared to the ABF estimate $G_t$.
As the ABF simulation progresses, $G_t$ converges towards a conservative vector field, and the ABF and pABF forces tend to become identical.
This is visible in the right column of Figure~\ref{fig:Helmholtz}, where the same quantities are plotted for a later point in the ABF simulation, much closer to convergence.
The pABF force at 3~ns (middle left), in the visited regions, is closer than the mean force (top left) to the converged mean force (top right).
Thus, within sampled regions, the conservative biasing force of pABF converges faster than the mean force of standard ABF.

\subsection{Rate of exploration and convergence of pABF}

\begin{figure}[!ht]
\centering
 \includegraphics[width=0.49\textwidth]{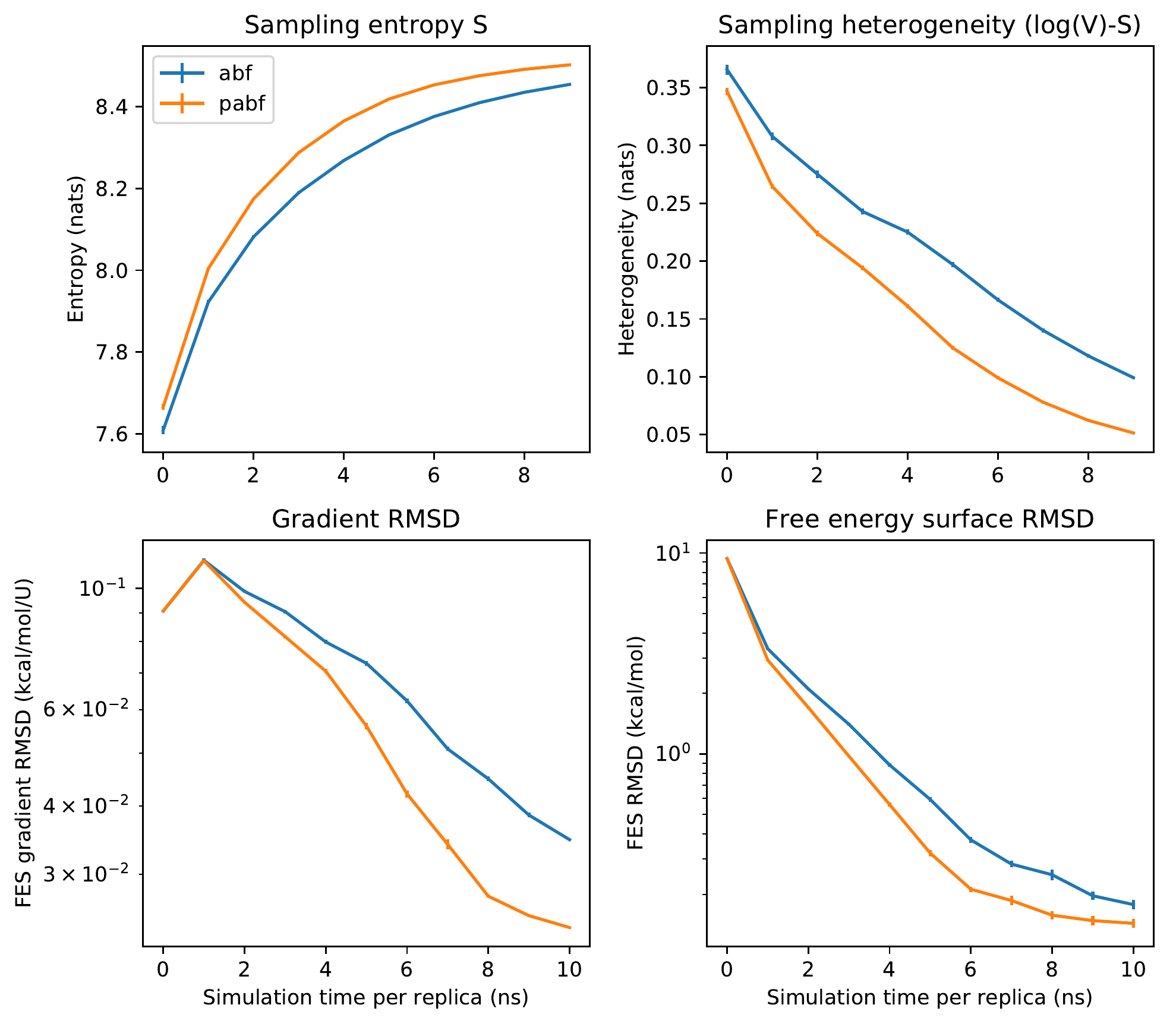}

 \caption{Exploration and convergence metrics for the 2d NANMA example (no hidden barriers), with and without pABF. }
\label{fig:pabf_nanma}
\end{figure}

Convergence of free energy calculations is usually monitored using relative metrics, measuring convergence towards a known target estimate of the free energy or its gradient, or in the absence of such a reference, towards the final estimate of the present calculation.
Convergence in terms of gradient is a more direct measure of the output of ABF, however it is sensitive to noise in the gradient estimates: therefore methods that yield less accurate but smoother gradient values fare better by that metric, even if the corresponding free energy is more biased.
Convergence in free energy, conversely, requires an arbitrary anchoring of the free energy surfaces.
Here the minimum value is set to zero before computing the RMSD.

To complement these relative metrics that depend on a known reference, I use absolute metrics of sampling in colvar space.
A pure exploration metric is the natural logarithm of the visited volume, measured as the number of bins with nonzero samples: $\log(V)\equiv\log(n_\mathrm{visited})$, ignoring the additive constant corresponding to the log of the bin volume.

Since ABF converges towards uniform sampling at long times, the heterogeneity of sampling can be used as an absolute convergence metric.
First the sampling entropy $S$ is calculated as:
\begin{equation}
 S = -\sum_{i=1}^{n_\mathrm{visited}} \frac{n_i}{N} \log\left( \frac{n_i}{N}\right)
\end{equation}
where $n_i$ is the number of samples collected in bin $i$ and $N = \sum_i n_i$, and the sum is calculated over visited bins only.
For a uniform distribution, $S = \log(n_\mathrm{visited})$.
Therefore, $h \equiv \log(n_\mathrm{visited}) - S \geq 0$, a value of 0 corresponding to uniform sampling.
I use the quantity $h$ as a measure of the heterogeneity of sampling.
These metrics are implemented in the Python module provided as Supplementary Material.

These convergence and exploration metrics for ABF on NANMA are shown in Figure~\ref{fig:pabf_nanma}.
ABF and pABF exhibit qualitatively similar exploration and convergence profiles, but pABF explores faster and more uniformly, and converges faster to the reference free energy profile and gradient.
Therefore, the reduced-variance biasing force of pABF is beneficial to convergence on the NANMA example.

% \begin{figure}[!ht]
% \centering
%  \includegraphics[width=0.49\textwidth]{Figures/nanma_3d_conv}
% 
%  \caption{Convergence and exploration metrics for a 3d free energy landscape on the NANMA example with and without pABF. }
% \label{fig:pabf_nanma_3d}
% \end{figure}

Since NANMA is intrinsically low-dimension, it is better viewed as a toy model than a relevant benchmark for sampling algorithms.
Previous work has shown the multimodal character of deca-alanine conformations in vacuum,\cite{Henin2010a} making it a minimal yet demanding system for conformational sampling.
The present ABF simulations use the first 3 eigenmodes from dihedral principal component analysis (dPCA)\cite{Altis2007}.

\begin{figure}[!ht] 
\centering
 \includegraphics[width=0.49\textwidth]{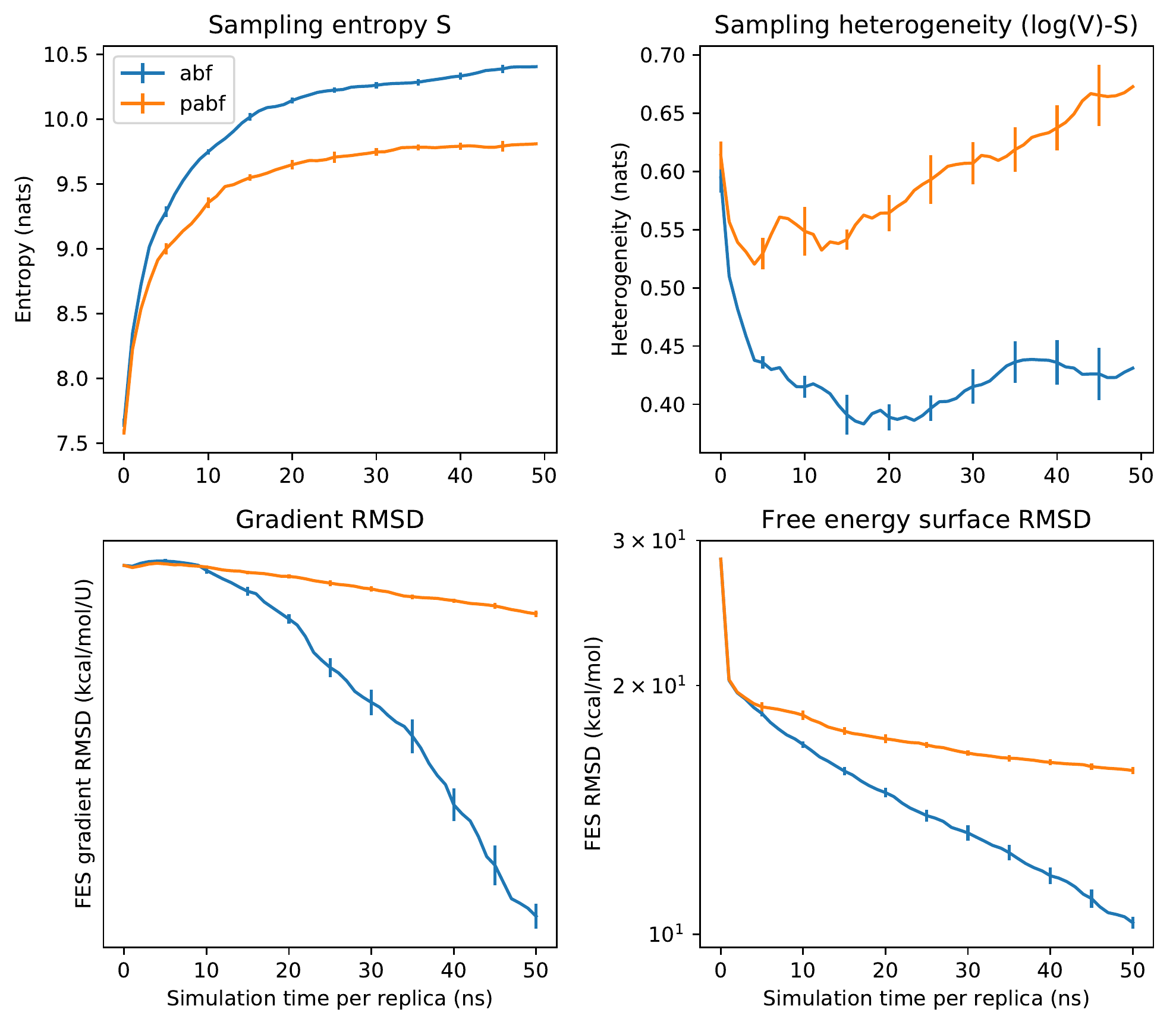}

 \caption{Exploration and convergence metrics for the 3D deca-alanine example (hidden barriers), with and without pABF. }
\label{fig:pabf_10ala}
\end{figure}
% \twocolumngrid

% pABF with fullSamples zero is much more stable than ABF (SI figure X), but yields more variance than fs 200 in the low-lying regions of the FES as detected by the KL divergence metric.

As in the previous example, ABF and pABF converge to the same 3d free energy landscape for deca-alanine (Figure~\ref{fig:pabf_10ala}), with a global RMSD smaller than $2.1\times 10^{-1}$~kcal/mol from the same reference, indicating a clear consensus.
In contrast with the convergence results of Figures~\ref{fig:Helmholtz} and \ref{fig:pabf_nanma}, however, pABF does not exhibit improved convergence over standard ABF, by any metric.
The salient feature of pABF dynamics on this system is that its exploration is considerably slower than standard ABF.
This is in sharp contrast with both the present results on the simpler NANMA example, and the results of Alrachid and Lelièvre on a model system. Those two systems share the property of possessing an intrinsically low-dimension slow space (2d), without significant ``hidden barriers'' along other degrees of freedom -- the ``orthogonal space''.\cite{Zheng2009,Paz2018}

An unexpected result is that reducing the variance of the biasing force does not always lead to faster convergence of ABF.
The presence of noise in the biasing force in the transient regime of ABF may enhance diffusion, leading to faster exploration.
It might explain the efficacy of ``classic'' metadynamics\cite{Laio2002} for exploration, with its ever-varying biasing potential that promotes constant motion in collective variable space.
``Well-tempered'' metadynamics\cite{Barducci2008} is a compromise between this initial behavior and long-term convergence of the free energy surface, which can also seen as switching from exploration to exploitation.
This is also exploited in the \textit{meta-eABF} and \textit{WTM-eABF} methods, combinations of metadynamics and ABF.\cite{Fu2019}
Another possibility is that classic ABF is more responsive than pABF to local fluctuations of the mean force corresponding to fluctuations in the orthogonal space,\cite{Zheng2008} leading to more efficient crossing of hidden barriers during the exploration phase.

pABF has common properties with ABF(GPR)\cite{Mones2016}: in ABF(GPR), the FES is reconstructed by Gaussian Process Regression, then differentiated to obtain the biasing force, which is constrained to be a true gradient, as the biasing force of pABF. ABF(GPR) was shown to explore the 2d FES of NANMA significantly faster than standard ABF, and approximately as fast as a constant bias exactly compensating the underlying FES.
However, the same study showed that the metadynamics leads to even faster exploration, presumably due to its self-avoiding behavior, regardless on the presence of free energy barriers, or lack thereof.

In principle, pABF can be combined with eABF, however, the mean force from eABF is already smoother than that from ABF\cite{Lesage2017}, suggesting that the variance reduction benefit could be lower than with standard ABF.

% \clearpage 
\section{Conclusion}

I have presented a flexible and efficient multidimensional integration algorithm, which is integrated in the Colvars Module implementations of ABF and eABF-CZAR, so that they yield accurate free energy surfaces in dimensions 2 and 3.
I have also implemented projected ABF, a reduced-variance version of ABF, together with new streamlined, efficient and flexible tools for visualization and analysis of ABF simulation results.
Free energy gradients in dimension higher than 3, which are rarely used, can still be integrated using the existing integration tool in arbitrary dimension.\cite{Henin2010a}

I have also investigated pABF, which had previously been derived theoretically, and tested numerically on a toy model that, crucially, contained a single slow degree of freedom, without hidden barriers.\cite{Alrachid2015}
The tests show that pABF is a proven lower-variance method, with surprisingly deleterious consequences on exploration.
These results are not sufficient to exclude a potential benefit of pABF in slower-relaxing systems, for which variance reduction and the associated slowdown in exploration could avoid long-lasting non-equilibrium biases in the free energy estimate.\cite{Miao2020}
However, it is possible that pABF reduces noise components with short autocorrelation times, not the long-lived biases due to slow orthogonal relaxation.
Other modifications of ABF tend to reduce the diffusive timescales, such as stratification or combinations with metadynamics.\cite{Fu2019}
However, metadynamics and related methods, while producing faster exploration of collective variable space, do not address orthogonal barriers.

Stratification can be used to limit the diffusion times by reducing the space available to each trajectory. This strategy is compatible with pABF.

Comparing different methods on equal footing remains difficult, even when they are as close as ABF and pABF, because all tunable parameters should be optimized for each method independently.
When some of the parameters play the role of adaptation rates, the presence of realistic relaxation, including a slow orthogonal space, is necessary to fully assess the real-world benefits of each method.

\section*{Acknowledgments}

The author gratefully acknowledges support from the Laboratoire International Associé CNRS-UIUC, and from the French National Research Agency under grant LABEX DYNAMO (ANR-11-LABX-0011).
The author thanks Dr. Tony Lelièvre for fruitful discussions, and Mr. Geoffrey Letessier for assistance with local computational resources.
The author is grateful to the referees of this work for their constructive comments.
Computational work was performed using HPC resources from LBT/HPC.

\section*{Supporting Information}

The Supporting Information document contains: a description of the software architecture of the Colvars Module; a summary of the availability of the software, independently and within parent software packages; an example script using the provided Python module for analysis of ABF data.

This information is available free of charge via the Internet at http://pubs.acs.org

\bibliography{references}

\newpage
\section*{For Table of Contents Only}

\begin{center}
 \includegraphics[width=0.7\textwidth]{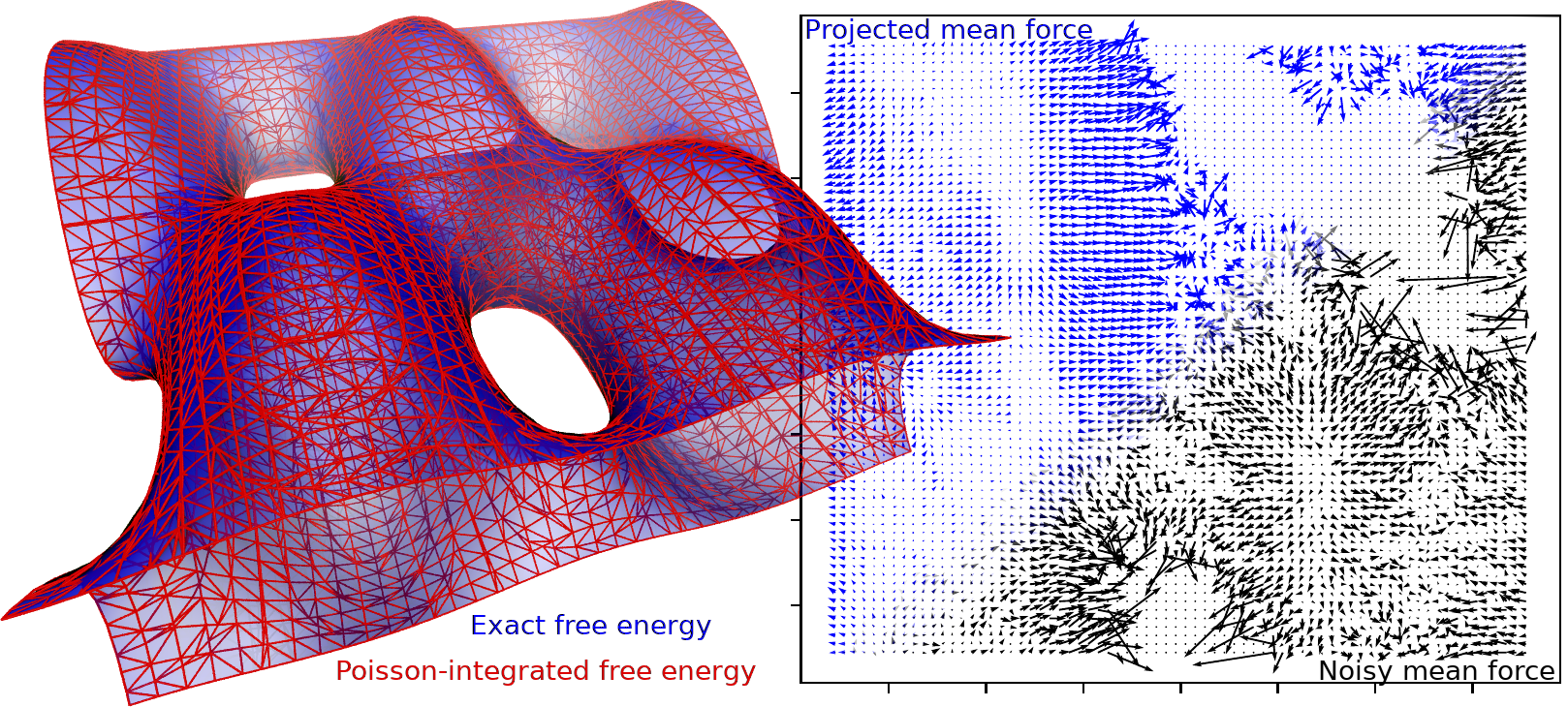}
\end{center}

\end{document}

\maketitle

\section{Architecture of the ABF implementation of the Colvars Module}

ABF is implemented as a specialization of the generic \texttt{colvarbias} class within the Colvars module\cite{Fiorin2013}, building mainly on the \texttt{colvar} class to access properties of user-defined collective variables, and \texttt{colvargrid} to store and manage discretized representations of the sampling histogram, free energy gradient, and free energy surface.
In particular, the derived class \texttt{integrate\_potential} implements Poisson integration as discussed here.
Although they are implemented in arbitrary dimension, the memory requirement for storing dense grids increases exponentially with dimension, practically limiting typical applications to dimension 1 to 3.
Free energy surfaces in higher dimension are best represented using sparse localized basis functions.\cite{Laio2002, Maragliano2008, Invernizzi2020}
Apart from the features described here, the implementation of ABF within Colvars has been discussed in detail elsewhere\cite{Henin2010a, Fiorin2013, Comer2015, Lesage2017}.

% \section{Citing methods and their software implementations}
% 
% Academic software development is only sustainable if it is duly acknowledged in publications that make use of said software. Table~\ref{tab:refs} lists the relevant references to be cited when making use of enhanced sampling methods in the Collective Variables Module.
% 
% % \onecolumngrid
% 
% \begin{table*}[h!]
%  \begin{tabular}{|c|c|c|}
%   \hline
%   \textbf{Algorithm} &
%  \textbf{Ref. for original algorithm} &
% \textbf{Ref. for Colvars}
% \\
% & &  \textbf{implementation}\\
%   \hline
% metadynamics&
%  Laio and Parrinello 2002 \cite{Laio2002}&
%  Fiorin et al. 2013 \cite{Fiorin2013}\\
% well-tempered metadynamics&
%  Barducci et al. 2008 \cite{Barducci2008}&
%  Fiorin et al. 2013 \cite{Fiorin2013}\\
% ABF&
%  Darve and Pohorille 2001 \cite{Darve2001}&
%  Hénin et al. 2010 \cite{Henin2010a}\\
% eABF&
%  Lelièvre et al. 2007 \cite{Lelievre2007}&
%  Lesage et al. 2017 \cite{Lesage2017}\\
% Poisson integration &
% LRS2010, Alrachid and Lelièvre 2015 \cite{Lelievre2010, Alrachid2015}&
%  this work\\
% pABF&
% LRS2010, Alrachid and Lelièvre 2015 \cite{Lelievre2010, Alrachid2015}&
%  this work\\
% Python analysis and plotting tools&
%  –&
%  this work\\
%  \hline
%  \end{tabular}
% \caption{Reference citations for the implementation of free energy algorithms in Colvars}
% \label{tab:refs}
% \end{table*}
% % \twocolumngrid

\section{Software availability}

\subsection{ABF with Poisson integration and pABF}
The source code is available on the official GitHub repository of the Colvars module:\\
\texttt{https://github.com/Colvars/colvars}
C++ sources are available for all back-ends: NAMD, \cite{Phillips2020} LAMMPS, \cite{Plimpton1995} Gromacs, \cite{Abraham2015} and VMD \cite{Humphrey1996} for trajectory analysis.
Binary distributions are available in all standard builds of the MD codes NAMD \cite{Phillips2020} and LAMMPS, \cite{Plimpton1995} as well as VMD.

\subsection{Analysis scripts}

Analysis tools for ABF simulations are provided as Supplementary Material, in the form of the Python module \texttt{colvars\_grid}, and an example Jupyter notebook built thereon.
These implement reading and writing of grid files written by Colvars, as well as the convergence and exploration metrics discussed below.
They are easily extensible by users with moderate proficiency in Python.
Updated versions of these files will be maintained in the Colvars repository on GitHub:\\
\texttt{https://github.com/Colvars/colvars/colvartools}.

The Python class \texttt{colvars\_grid} is provided to streamline the analysis and plotting of ABF data. An example script using that class to plot multiple graphs is given below:
\begin{verbatim}
import colvars_grid as cvg
import matplotlib.pyplot as plt

# 2d gradients as arrows
grad = cvg.colvars_grid('mysim.grad')
plt.quiver(*grad.meshgrid(), *grad.data)

# 2d PMF as level lines
pmf = cvg.colvars_grid('mysim.pmf')
plt.contour(*pmf.meshgrid(), *pmf.data)

# 3d surface plot for a 2d free energy surface
# (PMF), with contour plot at z=0
from mpl_toolkits.mplot3d import Axes3D
pmf = cvg.colvars_grid('mysim.pmf')
fig = plt.figure()
ax = fig.add_subplot(111, projection='3d')
ax.plot_surface(*pmf.meshgrid(), pmf.as_array())
ax.contour(*pmf.meshgrid(), pmf.as_array(),
zdir='z', offset=0)

# PMF convergence plot
pmf_traj = cvg.colvars_grid('mysim.hist.pmf')
plt.plot(plf_traj.convergence())
\end{verbatim}

\bibliographystyle{unsrt}
\bibliography{database}